\documentstyle [aps,preprint,epic,eepic,axodraw]{revtex}
\tightenlines
\begin{document}
\def\fnote#1#2{
\begingroup\def\thefootnote{#1}\footnote{#2}\addtocounter{footnote}{-1}
\endgroup}
\def\dslash{\not{\hbox{\kern-2pt $\partial$}}}
\def\eslash{\not{\hbox{\kern-2pt $\epsilon$}}}
\def\Dslash{\not{\hbox{\kern-4pt $D$}}}
\def\Aslash{\not{\hbox{\kern-4pt $A$}}}
\def\Qslash{\not{\hbox{\kern-4pt $Q$}}}
\def\Wslash{\not{\hbox{\kern-4pt $W$}}}
\def\pslash{\not{\hbox{\kern-2.3pt $p$}}}
\def\kslash{\not{\hbox{\kern-2.3pt $k$}}}
\def\qslash{\not{\hbox{\kern-2.3pt $q$}}}
\def\np#1{{\sl Nucl.~Phys.~\bf B#1}}
\def\pl#1{{\sl Phys.~Lett.~\bf B#1}}
\def\pr#1{{\sl Phys.~Rev.~\bf D#1}}
\def\prl#1{{\sl Phys.~Rev.~Lett.~\bf #1}}
\def\cpc#1{{\sl Comp.~Phys.~Comm.~\bf #1}}
\def\anp#1{{\sl Ann.~Phys.~(NY) \bf #1}}
\def\etal{{\em et al.}}
\def\half{{\textstyle{1\over2}}}
\def\be{\begin{equation}}
\def\ee{\end{equation}}
\def\ba{\begin{array}}
\def\ea{\end{array}}
\def\tr{{\rm tr}}
\title{Gluon interference effects in t\=t production at the Next
Linear Collider
\thanks{Research
supported in part by the DoE under grant DE--FG05--91ER40627.}}
\author{
%Linda Arvin
%\fnote{\dagger}{\tt linda@panacea.phys.utk.edu} and 
George Siopsis
\fnote{\dagger}{\tt gsiopsis@utk.edu}}
\address{Department of Physics and Astronomy, \\
The University of Tennessee, Knoxville, TN 37996--1200.\\
}
\date{November 1997}
\preprint{UTHEP--97--0901}
\maketitle
\begin{abstract}
We study the effects of gluon interference in the production and semi-leptonic
decay of a t\=t pair above threshold at the Next Linear Collider (NLC).
We calculate all matrix elements to next-to-leading order and use the resulting
expressions for the development of a Monte Carlo event generator.
Our results show effects at the level of 10\% in differential cross-sections.
We thus extend previous results obtained by analytical
means in the soft-gluon limit.
\end{abstract}
\renewcommand\thepage{}\newpage\pagenumbering{arabic}

\section{Introduction}

The discovery of the top quark~\cite{ref1,ref2} has opened up a new avenue for the exploration of new physics.
The large mass of the top ($m_t = 176\pm 8\pm 10$ GeV (CDF)~\cite{ref1};
$m_t = 199 {+ 19\atop -21} \pm 22$ GeV (D0)~\cite{ref2})
and, consequently, its large decay width ($\Gamma_t = 1.57$ GeV~\cite{ref3}), permit the use of
perturbation theory, because the top quark is expected to decay before it has a chance to
hadronize. It is unlike the other (lighter) quarks, which form resonances that are outside
the realm of perturbation theory. It is therefore possible to perform a precision analysis
of top quark production and decay, which facilitates the identification of effects attributable
to new physics beyond the Standard Model~\cite{ref4}.

The best laboratory for such explorations is an $e^+e^-$ collider, because the initial state
is free of color effects~\cite{ref5}. Moreover, it is advantageous to have polarized beams for a more efficient
study of the electroweak properties of the top quark.
Such a machine (the Next Linear Collider (NLC)) will hopefully be
built in the not-so-distant future. Detailed studies of processes through which one may extract
useful information have already been performed~\cite{ref5a}. Gluon radiation contributes at the 10\% level,
in general. Although it does not modify inclusive cross-sections, it has a non-negligible effect
on various final-state distributions. Therefore, it has to be included in a precision study of
top quark physics. In studying gluon radiation, one cannot ignore the effects of gluon interference.
The latter depends upon the value of $\Gamma_t$ and may be an important tool in the study of
this parameter, in addition to precision studies~\cite{ref6}.

The decay width of the top quark is saturated by the decay mode $t\to W^+b$, because in the
Standard Model $|V_{tb}| \approx 1$. The $W$ boson, in turn, may decay into quarks or leptons.
Therefore, there are four possible decay modes in t\=t production and decay. The branching
ratio for the channel in which both $W$ bosons decay into leptons is about 10\%. This is the
channel we study here, because it gives the cleanest signal. We shall calculate all amplitudes
analytically, with the aid of the symbolic manipulation package FORM~\cite{ref7}. The resulting expressions
will then be used for the development of a Monte Carlo event generator. The generator is based on
the C++ program written by Schmidt for the study of the same process~\cite{ref8},
\begin{equation}
  \label{eq0}
  e^+e^- \to \gamma^\star \;,\; Z^\star \to t\bar t \to W^+b\; W^-\bar b \to
\ell^+\nu b \; \ell^-\bar\nu\bar b
\end{equation}
We have added gluon interference effects to the generator. The results from our simulations are in
agreement with analytical results obtained by Khoze, {\em et al.}~\cite{ref9} in the soft gluon regime. They are also similar
to the results obtained by Peter and Sumino~\cite{ref10} in the threshold region, where color Coulomb effects dominate~\cite{ref11}.

Our paper is organized as follows. In Section~\ref{sec1}, we outline the tree-level analysis in order
to introdce the basic tools and establish the notation. In Section~\ref{sec2}, we discuss loop
amplitudes at $o(\alpha_s)$. In Section~\ref{sec3}, we introduce real gluons and compute
the amplitudes for gluon radiation. 
In Section~\ref{sec4}, we discuss the development of the Monte Carlo event generator based
on the analyical results obtained in the previous sections. We also present numerical results.
Finally, in Section~\ref{sec5}, we present conclusions and discuss future directions.
%Top decay and branching ratios. Why have we chosen the small one?
%compare with Khoze. MC based on Schmidt.

%Summary of sections

\section{Tree-level analysis}
\label{sec1}

\begin{figure}
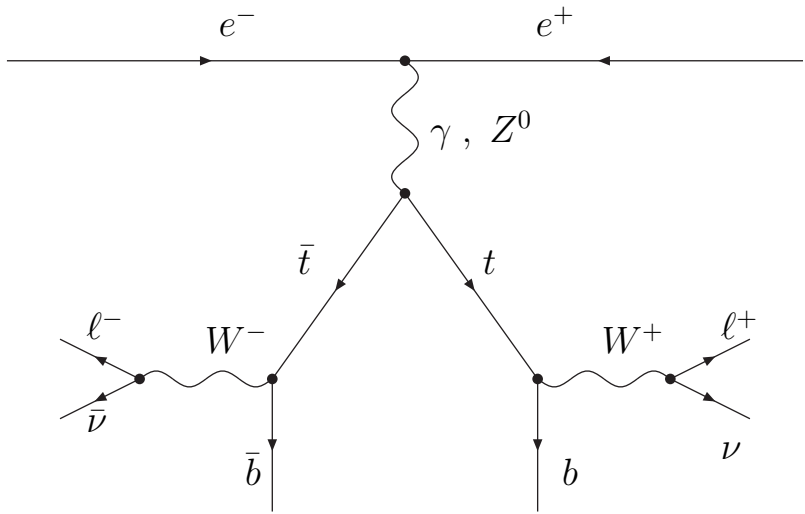


\vspace{3in}
\caption{t\=t production and decay at tree level.}
\label{fig1}
\end{figure}

At tree level~(Fig.~\ref{fig1}), the amplitude for t\=t production and decay factorizes.
With polarized $e^+e^-$ beams, one can perform a clean helicity analysis,
because the spins of the top and anti-top quarks get transferred to the final
products. It is then advantageous to express the amplitudes in terms of helicity
states~\cite{ref8}. To this end, one defines form factors for the
production and decay stages.

The form factor for t\=t production is the
neutral-current coupling ($\gamma$-t-\=t and Z-t-\=t vertices)
\begin{equation}
  \label{eq1}
  {\cal M}_{\gamma,Z}^\mu = e\gamma^\mu (F_{\gamma,Z}^{1V} +
F_{\gamma,Z}^{1A} \gamma^5) + {e\over 2m_t} \sigma^{\mu\nu} k_\nu
(F_{\gamma,Z}^{2V} + F_{\gamma,Z}^{2A} \gamma^5)\;,
\end{equation}
where in the Standard Model the only non-vanishing coupling constants at
tree level are
\begin{equation}
  \label{eq2}
F_\gamma^{1V} = Q_t = {2\over 3}\;,\quad
F_Z^{1V} = {1\over 2\sin (2\theta_W)} - {2\over 3} \tan\theta_W
\;,\quad F_Z^{1A} = - {1\over 2\sin (2\theta_W)}\;.
\end{equation}
The magnetic dipole coupling ($F^{2V}$) receives loop corrections at
$o(\alpha_s)$. The electric dipole coupling ($F^{2A}$) is CP violating
and remains zero to $o(\alpha_s^2)$.

The $t\to W^+b$ decay vertex is described by the form factor
(top-charged-current coupling)
\begin{equation}
  \label{eq3}
{\cal M}_W^\mu = {g\over\sqrt 2} \gamma^\mu (F_W^{1L} P_L +
F_W^{1R} P_R) + {ig\over 2\sqrt 2 m_t} \sigma^{\mu\nu} k_\nu
(F_W^{2L} P_L + F_W^{2R} P_R)\;,
\end{equation}
where $P_{R,L} = {1\over 2} (1\pm\gamma^5)$, and similarly
for the $\bar t\to W^-\bar b$ decay vertex. In the Standard Model
at tree level, $F_W^{1L} = 1$ and all other coupling constants vanish.

Finally, the $W^+\to \ell^+\nu$ decay is described by a left-handed
current coupling of strength ${igm_W\over \sqrt 2}$, and similarly for
$W^-\to \ell^-\bar\nu$.

Scattering amplitudes can be conveniently expressed in terms of helicity
angles~\cite{ref8}. There is a set of helicity angles for each vertex.
Different sets are defined with respect to different Lorentz frames. This
assumes the factorizability of the amplitudes. Since we want to study gluon
interference, we have to consider non-factorizable amplitudes at one-loop
level. Therefore, we will use a single Lorentz frame (laboratory frame).
To simplify the expressions, we introduce polarization vectors for the
$W^\pm$ bosons defined with respect to the b (\=b) quarks (we set
$m_b = 0$ in all matrix elements), respectively,
$$\left( W_L^+ \right)^\mu = {1\over 4\sqrt 2 \; p_{W^+}\cdot p_b}\;
\langle b_L| \; [\; \gamma^\mu \;,\; \pslash_{W^+} \; ] \;
|b_L\rangle\quad, \quad W_R^+ = W_L^{+\ast}$$
\begin{equation}
  \label{eq4}
\left( W_Z^+ \right)^\mu = {1\over m_W} \; p_{W^+}^\mu
- {m_W\over p_{W^+}\cdot p_b} \; p_b^\mu
\end{equation}
and similarly for $(W^-)^\mu$ (only one transverse polarization of each $W^\pm$
contributes to matrix elements).
Longitudinally polarized $W$ bosons are very important because the study of
longitudinal modes probes the electroweak symmetry-breaking mechanism.

The tree-level amplitudes can be written
\begin{equation}
  \label{eq5}
{\cal T}_{\gamma,Z} (L, W^-, W^+) = \left(
F_{\gamma,Z}^L F_{\{\gamma,Z\} t\bar t}^L
{\cal T}^{(1)} (L, W^-, W^+) - m_t^2F_{\gamma,Z}^L
F_{\{\gamma,Z\} t\bar t}^R {\cal T}^{(2)} (L, W^-, W^+)
\right) {\cal W}^+ {\cal W}^-
\end{equation}
where we chose left-handed polarization for the electron beam, for
definiteness.
$F_{\gamma,Z}^L$ and $F_{\{\gamma,Z\} t\bar t}^R$ are
appropriate form factors given in terms of the form factors in
Eqs.~(\ref{eq1}), (\ref{eq2}), and (\ref{eq3}), and
$${\cal T}^{(1)} (L, W^-, W^+) = \langle \bar b_R| 
\Wslash^- \pslash_{\bar t}| e_R^+\rangle
\langle e_L^-|\pslash_t\Wslash^{+\ast}|b_L\rangle$$
\begin{equation}
  \label{eq5a}
{\cal T}^{(2)} (L, W^-, W^+) = \langle e_R^+| 
\Wslash^{+\ast} | b_L\rangle
\langle \bar b_R|\Wslash^-|e_L^-\rangle
\end{equation}
The factors
\begin{equation}
  \label{eq5b}
{\cal W}^+ = \langle \ell_R^+ | \Wslash^+ |\nu_L \rangle
\;,\;
{\cal W}^- = \langle \bar\nu_R | \Wslash^{-\ast}| \ell_L^- \rangle
\end{equation}
represent the leptonic decays of the $W^\pm$ bosons, respectively.
The differential cross-section for left-handed polarized beams at tree-level is
\begin{equation}
  \label{eq6}
d \sigma_{tree} (L)= \left|\sum \Big({\cal T}_\gamma
(L, W^-, W^+) +
{\cal T}_Z (L, W^-, W^+)\Big)\right|^2
\end{equation}
where we sum over the polarizations of $W^\pm$.

\section{Virtual gluons}
\label{sec2}

\subsection{General}

\begin{figure}
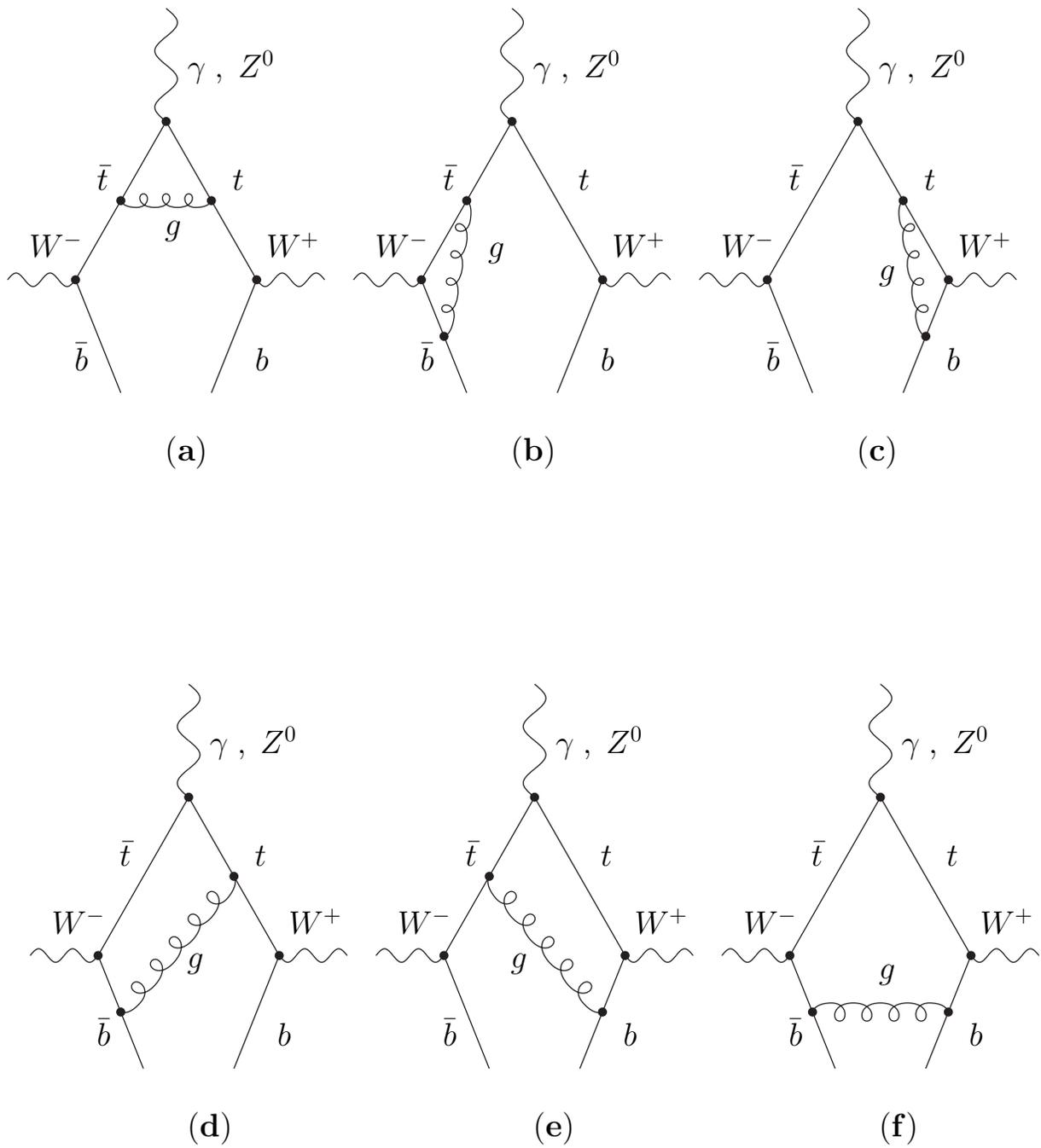

 %
\vspace{5in}
\caption{t\=t production and decay at one-loop level.}
\label{fig2}
\end{figure}

Loop diagrams at $o(\alpha_s)$ include a pentagon diagram (gluon exchange
between the b and \=b quarks; see Fig.~\ref{fig2}{\em (f)}), which renders the calculation of
$o(\alpha_s)$ effects cumbersome. To systematically calculate these
effects, observe that all loop diagrams contain one or more internal t or \=t
legs, which are rapidly decaying quarks. To take advantage of this fact, we
define the quantities
\begin{equation}
  \label{eq7}
\eta_1 = (p_{\bar t}^2 - m_t^2)/m_t^2 \;,\quad
\eta_2 = (p_t^2 - m_t^2)/m_t^2 \;.
\end{equation}
Due to the presense of propagators
\begin{equation}
  \label{eq8}
{1\over \pslash - m_t - i\Gamma_t/2}\;,
\end{equation}
where $p^\mu$ is the momentum of t (\=t), the integral over phase space
contains factors
\begin{equation}
  \label{eq9}
\int {d\eta_1 \over \eta_1^2 + \Gamma_t^2/m_t^2}
\int {d\eta_2 \over \eta_2^2 + \Gamma_t^2/m_t^2}\;,
\end{equation}
showing that $\eta_1$ and $\eta_2$ are small parameters,
$o(\Gamma_t/m_t)$. Physically, this means that t and \=t are produced
nearly on shell.

The rest of the amplitude is a function of $\eta_1$ and $\eta_2$.
Since $\eta_1$ and $\eta_2$ are small, it is advantageous to expand the
amplitude in these parameters,
\begin{equation}
  \label{eq10}
{\cal A} (\eta_1,\eta_2) = {\cal A}\Big|_{\eta_1=\eta_2=0}
+\eta_1 \left. {\partial {\cal A}\over \partial\eta_1}
\right|_{\eta_2=0} + \eta_2 \left. {\partial {\cal A}\over
\partial \eta_2} \right|_{\eta_1=0} + o(\eta_{1,2}^2)\;.
\end{equation}
where ${\cal A}$ needs to include both virtual and real gluon
contributions (soft and collinear), in order to avoid singularities,
\begin{equation}
  \label{eq10a}
{\cal A} = {\cal A}^{virtual} + {\cal A}^{soft} + {\cal A}^{collinear}
\end{equation}
The first term in Eq.~(\ref{eq10})
is the set of triangle diagrams (Fig.~\ref{fig2}{\em (a-c)}). This set of diagrams is
gauge-invariant if we set $\eta_1=\eta_2=0$. They represent vertex
(form-factor) corrections~\cite{ref8}. The rest of the terms represent
gluon interference. To calculate them, notice that if we set $\eta_2 = 0$,
then we are placing the top quark on shell, thus introducing a cut.
The diagrams contributing in this case are those containing
a gluon which does not cross the cut (Fig.~\ref{fig2}{\em (a,b,d)} and
Fig.~\ref{fig3}{\em (a-c)}). This set is gauge-invariant at the
point $\eta_2 = 0$. Notice that the pentagon diagram (Fig.~\ref{fig2}{\em (f)})
is not included in
this set. Their contribution to the amplitude is
\begin{equation}
  \label{eq11}
{\cal A}_1 = {\cal A}\Big|_{\eta_2 = 0} = {\cal A}
\Big|_{\eta_1 = \eta_2 = 0} + \eta_1 \left. {\partial {\cal A}
\over \partial\eta_1} \right|_{\eta_2=0} + o(\eta_{1,2}^2)\;.
\end{equation}
Similarly, if we set $\eta_1 = 0$ by cutting the \=t line,
we obtain a set of diagrams which is gauge-invariant at the
point $\eta_1 = 0$ (Fig.~\ref{fig2}{\em (a,c,e)} and
Fig.~\ref{fig3}{\em (b-d)}). Their contribution to the amplitude is
\begin{equation}
  \label{eq12}
{\cal A}_2 = {\cal A}\Big|_{\eta_1 = 0} = {\cal A}
\Big|_{\eta_1 = \eta_2 = 0} + \eta_2 \left. {\partial {\cal A}
\over \partial\eta_2} \right|_{\eta_1=0} + o(\eta_{1,2}^2)\;.
\end{equation}
The pentagon diagram (Fig.~\ref{fig2}{\em (f)})
contributes to $o(\eta_{1,2}^2)$.

The $o(\alpha_s)$ amplitude (Eq.~(\ref{eq10})) can thus be written
\begin{equation}
  \label{eq13}
{\cal A} (\eta_1,\eta_2) = {\cal A} (0,0) + ({\cal A}_1 - {\cal A}
(0,0)) + ({\cal A}_2 - {\cal A} (0,0)) + o(\eta_{1,2}^2)\;.
\end{equation}
We shall calculate each term separately.

\subsection{The limit of on-shell t and \=t}

In this case, the amplitude factorizes. There are three triangles
(Fig.~\ref{fig3}{\em (a-c)})
contributing and have already been calculated~\cite{ref8}. Here, we express
them in terms of $W^\pm$ polarization vectors~(Eq.~(\ref{eq4})), mainly for
completeness. For definiteness, we present results for left-handed beam
polarization. Right-handed polarization can be obtained by interchanging
$e^+$ and $e^-$.

The $o(\alpha_s)$ correction to the t\=t production vertex is
\begin{equation}
  \label{eq14}
C_F (g\mu^\epsilon)^2
\int {d^{4-2\epsilon}k\over (2\pi)^{4-2\epsilon}} \; {{\cal C}_{\{\gamma,Z\} t\bar t}
(L, W^-, W^+) \over k^2 (k^2 -2k\cdot p_{\bar t})(k^2 +2k\cdot p_t)}
\end{equation}
where
\begin{equation}
  \label{eq14a}
{\cal C}_{\{\gamma,Z\} t\bar t} (L, W^-, W^+) = F_{\gamma,Z}^L F_{\{\gamma,Z\} t\bar t}^L
{\cal C}_{t\bar t}^{(1)} (L, W^-, W^+) - m_t^2F_{\gamma,Z}^L F_{\{\gamma,Z\} t\bar t}^R {\cal C}_{t\bar t}^{(2)} (L, W^-, W^+)
\end{equation}
$${\cal C}_{t\bar t}^{(1)} (L, W^-, W^+) = 2p_t\cdot p_{\bar t}\langle e_L^-| \pslash_t \Wslash^{+\ast} |b_L\rangle
\langle \bar b_R| \Wslash^- \pslash_{\bar t} | e_R^+ \rangle$$
$$+ \langle e_L^- | \pslash_{\bar t}\pslash_t\kslash\Wslash^{+\ast} |b_L\rangle
\langle \bar b_R| \Wslash^- \pslash_{\bar t} |e_R^+ \rangle$$
$$-\langle e_L^-| \pslash_t \Wslash^{+\ast} |b_L\rangle \langle \bar b_R| \Wslash^- \pslash_{\bar t}
\pslash_t\kslash |e_R^+\rangle$$
\begin{equation}
  \label{eq14b}
- \langle e_R^+| \kslash\pslash_t \Wslash^{+\ast} |b_L\rangle
\langle \bar b_R| \Wslash^- \pslash_{\bar t}\kslash |e_L^-\rangle
\end{equation}
$${\cal C}_{t\bar t}^{(2)} (L, W^-, W^+) = -2p_t\cdot p_{\bar t}\langle e_R^+ | \Wslash^{+\ast} |b_L\rangle
\langle \bar b_R| \Wslash^- |e_L^- \rangle$$
$$-\langle e_R^+ | \kslash\pslash_{\bar t}\Wslash^{+\ast} |b_L\rangle \langle \bar b_R| \Wslash^- |e_L^- \rangle$$
$$+\langle e_R^+ | \Wslash^{+\ast} |b_L\rangle \langle \bar b_R| \Wslash^-\pslash_t\kslash|e_L^- \rangle$$
\begin{equation}
  \label{eq14c}
+\langle e_L^-| \kslash\Wslash^{+\ast} |b_L\rangle \langle \bar b_R| \Wslash^- \kslash
|e_R^+\rangle
\end{equation}
These expressions can be further simplified by employing helicity states for the on-shell t and \= t quarks. We shall not do it here,
because we are interested in the more general case of off-shell quarks.
Notice that we obtain the correct infrared limit (after we subtract ultraviolet divergences)
\begin{equation}
  \label{eq14d}
C_F (g\mu^\epsilon)^2
\int {d^{4-2\epsilon}k\over (2\pi)^{4-2\epsilon}} \; {1\over k^2}
\left({-p_{\bar t}^\mu + k^\mu /2\over k^2 -2k\cdot p_{\bar t}} - {p_t^\mu + k^\mu /2
\over k^2 +2k\cdot p_t}\right)^2\; {\cal T}_{\gamma,Z} (L, W^-, W^+) 
\end{equation}
The $o(\alpha_s)$ correction to the $\bar t \to W^-\bar b$ decay vertex is
\begin{equation}
  \label{eq15}
C_F (g\mu^\epsilon)^2
\int {d^{4-2\epsilon}k\over (2\pi)^{4-2\epsilon}} \; {{\cal C}_{\{\gamma,Z\} \bar t} (L, W^-, W^+) \over k^2 (k^2 -2k\cdot p_{\bar b})(k^2 -2k\cdot p_{\bar t})}
\end{equation}
where
\begin{equation}
  \label{eq15a}
{\cal C}_{\{\gamma,Z\} \bar t} (L, W^-, W^+) = F_{\gamma,Z}^L F_{\{\gamma,Z\} t\bar t}^L
{\cal C}_{\bar t}^{(1)} (L, W^-, W^+) - m_t^2F_{\gamma,Z}^L F_{\{\gamma,Z\} t\bar t}^R {\cal C}_{\bar t}^{(2)} (L, W^-, W^+)
\end{equation}
\begin{equation}
  \label{eq15b}
{\cal C}_{\bar t}^{(1)} (L, W^-, W^+) = \langle e_L^-| \pslash_t \Wslash^{+\ast} |b_L\rangle
\langle \bar b_R| {\cal I} \pslash_{\bar t}|e_R^+\rangle
\end{equation}
\begin{equation}
  \label{eq15c}
{\cal C}_{\bar t}^{(2)} (L, W^-, W^+) = \langle \bar b_R| {\cal I} |e_L^-\rangle\langle e_R^+| \Wslash^{+\ast} |b_L\rangle
\end{equation}
\begin{equation}
  \label{eq15d}
{\cal I} = 2(-p_{\bar t}+k)\cdot p_{\bar b}\Wslash^- 
 - 2p_{\bar b}\cdot
W^-\kslash+\pslash_{\bar t}\kslash\Wslash^- +\kslash\Wslash^-\kslash
\end{equation}
Again, the correct infrared limit is obtained, similar to Eq.~(\ref{eq14d}).
The $o(\alpha_s)$ correction to the $\bar t\to \bar bW^-$ decay
vertex can be obtained from the $t\to bW^+$ decay vertex by conjugation.

\subsection{The limit of on-shell t}

Here we calculate the virtual gluon contribution to ${\cal A}_1$ (Eq.~(\ref{eq11})).
The vertex corrections we calculated above need to be modified to include the $o(\eta_1)$ effects
of an off-shell \= t quark.
We obtain additional contributions (Eqs.~(\ref{eq14b}), (\ref{eq14c}), (\ref{eq15b}) and (\ref{eq15c}))
$$\Delta {\cal C}_{t\bar t}^{(1)} (L, W^-, W^+) = m_t^2\eta_1\langle e_L^-| \pslash_t \Wslash^{+\ast} |b_L\rangle \langle \bar b_R| \Wslash^- 
(\pslash_t+\kslash) |e_R^+\rangle$$
\begin{equation}
  \label{eq16}
-m_t^2\eta_1 \langle e_R^+| \kslash\pslash_t\Wslash^{+\ast} |b_L\rangle
\langle \bar b_R| \Wslash^- |e_L^-\rangle
\end{equation}
\begin{equation}
  \label{eq17}
\Delta {\cal C}_{\bar t}^{(1)} (L, W^-, W^+) = 2
(-p_{\bar b}+k)\cdot W^-\; m_t^2\eta_1\langle e_L^-| \pslash_t \Wslash^{+\ast} |b_L\rangle
\langle \bar b_R
|e_R^+\rangle
\end{equation}
and $\Delta {\cal C}_{t\bar t}^{(2)} (L, W^-, W^+) = \Delta {\cal C}_{\bar t}^{(2)} (L, W^-, W^+) = 0$.

In addition to corrections to the triangle diagrams of $o(\eta_1)$,
we have to consider the box diagram in which a gluon is exchanged between \=b and the top quark (Fig.~\ref{fig2}{\em (d)}).
Its contribution is formally of $o(\eta_1)$,
\begin{equation}
  \label{eq18}
C_F (g\mu^\epsilon)^2
\eta_1 \int {d^{4-2\epsilon}k\over (2\pi)^{4-2\epsilon}} \; {{\cal D}_{\gamma,Z} (L, W^-, W^+) \over k^2 (k^2 -2k\cdot p_{\bar b})(k^2 -2k\cdot p_{\bar t}+m_t^2\eta_1)(k^2 +2k\cdot p_t)}
\end{equation}
where
\begin{equation}
  \label{eq18a}
{\cal D}_{\gamma,Z} (L, W^-, W^+) = F_{\gamma,Z}^L F_{\{\gamma,Z\} t\bar t}^L
{\cal D}^{(1)} (L, W^-, W^+) - m_t^2F_{\gamma,Z}^L F_{\{\gamma,Z\} t\bar t}^R {\cal D}^{(2)} (L, W^-, W^+)
\end{equation}
$${\cal D}^{(1)} (L, W^-, W^+) = \langle e_L^-|\pslash_t \Wslash^{+\ast} |b_L\rangle
\langle \bar b_R| (\pslash_t+\kslash)(-\pslash_{\bar b}+\kslash)\Wslash^-(- \pslash_{\bar t}+\kslash)|e_R^+\rangle$$
\begin{equation}
  \label{eq18b}
-\langle e_L| (-\pslash_{\bar b}+\kslash)\Wslash^-(-\pslash_{\bar t}+\kslash)|e_R^+\rangle
\langle \bar b_R|\kslash\pslash_t\Wslash^{+\ast}|b_L\rangle
\end{equation}
$${\cal D}^{(2)} (L, W^-, W^+) = \langle e_R^+|\Wslash^{+\ast}|b_L\rangle
\langle \bar b_R|\pslash_t (-\pslash_{\bar b}+\kslash) \Wslash^- |e_L^-\rangle$$
\begin{equation}
  \label{eq18c}
+ \langle e_R^+|\kslash(-\pslash_{\bar b}+\kslash) \Wslash^- |e_L^-\rangle
\langle \bar b_R|\Wslash^{+\ast}|b_L\rangle
\end{equation}
To perform the integral over the loop momentum in Eq.~(\ref{eq18}), we can use a standard reduction procedure
implemented via the symbolic manipulation program FORM~\cite{ref7}. The only cumbersome terms are those
with three loop-momentum insertions in Eq.~(\ref{eq18b}),
\begin{equation}
  \label{eq19}
\langle e_L^-|\pslash_t \Wslash^+ |b_L\rangle
\langle \bar b_R| \kslash\kslash\Wslash^-\kslash|e_R^+\rangle
-\langle e_L^-| \kslash\Wslash^-\kslash|e_R^+\rangle
\langle \bar b_R|\kslash\pslash_t\Wslash^+|b_L\rangle
\end{equation}
The first term in Eq.~(\ref{eq19}) is readily reduced to a triangle diagram, because $\kslash\kslash = k^2$
cancels one of the propagators in Eq.~(\ref{eq18}). To reduce the second term, we note that
\begin{equation}
  \label{eq20}
\kslash\Wslash^-\kslash = 2 k\cdot W^- \kslash - k^2 \Wslash^-
\end{equation}
For transverse polarization, we have (Eq.~(\ref{eq4}))
\begin{equation}
  \label{eq21}
k\cdot W^- = {1\over 4\sqrt 2 p_{W^-} \cdot p_{\bar b}}
\langle \bar b_L|\; [\; \kslash\;, \;\pslash_{\bar t} \; ]\; |\bar b_L\rangle
\end{equation}
When multiplied by $\langle \bar b_R|\kslash\dots$, we obtain an expression of the form
\begin{equation}
  \label{eq22}
\dots \kslash \pslash_{\bar b} \kslash\dots = \dots (2k\cdot p_{\bar b} \kslash
- k^2 \pslash_{\bar b} )\dots = \dots \Big( -\kslash (k^2 - 2k\cdot p_{\bar b})
+ (-\pslash_{\bar b} + \kslash) k^2 \Big)\dots
\end{equation}
which only contributes to triangles. For longitudinal $W^-$ polarization (Eq.~(\ref{eq4})),
\begin{equation}
  \label{eq23}
k\cdot W^- = {1\over m_W} \left( k\cdot p_{\bar t} - {p_{W^-} \cdot p_{\bar t}
\over p_{W^-} \cdot p_{\bar t}} k\cdot p_{\bar b} \right)
\end{equation}
which leads to triangles and a box with one less loop momentum insertions.
This is the advantage of introducing polarization vectors for $W^\pm$.

\subsection{The limit of on-shell \=t}

The amplitudes contributing to ${\cal A}_2$ (Eq.~(\ref{eq12})), which include
$o(\eta_2)$ effects, can be obtain from the expressions we derived above for ${\cal A}_1$
by conjugation.

\subsection{Both t and \=t off shell}

In the case where both t and \=t are off shell, we need to introduce corrections to the diagrams we considered
above, as well as the pentagon diagram with a gluon exchange between the b and \=b quarks (Fig.~\ref{fig2}{\em (f)}). These
additional contributions to the amplitude are of $o(\eta_{1,2}^2)$ and will not be considered here.
We shall only point out that they do not present additional computational difficulties. They can be reduced to
scalar amplitudes in the same manner.

\section{Real gluons}
\label{sec3}

The calculation of amplitudes for real gluon emission (Fig.~\ref{fig3})
follows the same lines as the calculation of the virtual
contributions. We introduce polarization vectors for the gluon defined with respect to the b or \=b quark,
accordingly ({\em cf.}~Eq.(\ref{eq4})),
\begin{equation}
  \label{eq24}
\epsilon_L^\mu = {1\over 4\sqrt 2 p_b\cdot k} \langle b_L| \; [ \gamma^\mu \;,\; \kslash \;]\;
| b_L \rangle\quad ,\quad
\epsilon_R = \epsilon_L^\ast
\end{equation}
where $k^\mu$ is the momentum of the gluon, and similarly for the \=b quark.
In the limit of on-shell t and \=t quarks and the narrow-width approximation,
one can clearly distinguish between three channels
of gluon production (t\=t production, t decay and \=t decay, respectively). These channels get blurred by
gluon interference effects. However, it is advantageous to keep them distinct from a programmatic viewpoint.
To this end, we split the phase space into three distinct regions.

\subsection{The \=t decay channel}

In this channel we define $p_{\bar t}^\mu = p_{W^-}^\mu + p_{\bar b}^\mu + k^\mu$ and
$p_t^\mu = p_{W^+}^\mu + p_b^\mu$. $\eta_{1,2}$ are still given by Eq.~(\ref{eq7}).
We set $\eta_2 = 0$ in the matrix elements, so that
the integral over soft and collinear gluons contributes to ${\cal A}_1$ (Eq.~\ref{eq11}).
The region of phase space corresponding to this channel is defined by
\begin{equation}
  \label{eq24a}
|\eta_1 m_t^2| < |( p_{\bar t} - k)^2 - m_t^2|
\end{equation}
There are three contributions to this channel; gluon emission from \=b, \=t and t (Fig.~\ref{fig3}{\em (a-c)}). They are,
respectively,
$$R_{\bar b} = {1\over -2k\cdot p_{\bar b}} \Big( F_\gamma^L F_{\gamma t\bar t}^L
{\cal R}_{\bar b}^{(1)} - m_t^2F_\gamma^L F_{\gamma t\bar t}^R {\cal R}_{\bar b}^{(2)}\Big)$$
$${\cal R}_{\bar b}^{(1)} = \langle \bar b_R| \eslash(\pslash_{\bar b}+\kslash)\Wslash^- \pslash_{\bar t}|
e_R^+\rangle
\langle e_L^-|\pslash_t\Wslash^{+\ast}|b_L\rangle$$
\begin{equation}
  \label{eq25}
{\cal R}_{\bar b}^{(2)} = \langle e_R^+| \Wslash^{+\ast} | b_L\rangle
\langle \bar b_R|\eslash(\pslash_{\bar b}+\kslash)\Wslash^-|e_L^-\rangle
\end{equation}
$$R_{\bar t} = {1\over -2k\cdot p_{\bar t} + m_t^2 \eta_1} \Big( F_\gamma^L F_{\gamma t\bar t}^L
{\cal R}_{\bar t}^{(1)} - m_t^2F_\gamma^L F_{\gamma t\bar t}^R {\cal R}_{\bar t}^{(2)}\Big)$$
$${\cal R}_{\bar t}^{(1)} = \langle \bar b_R|\Wslash^- \{(-\pslash_{\bar t}+\kslash)
\eslash\pslash_{\bar t}-m_t^2\eslash\}| e_R^+\rangle
\langle e_L^-|\pslash_t\Wslash^{+\ast}|b_L\rangle$$
\begin{equation}
  \label{eq26}
{\cal R}_{\bar t}^{(2)} = \langle e_R^+| \Wslash^{+\ast} | b_L\rangle
\langle \bar b_R|\Wslash^-\{- 2\epsilon\cdot p_{\bar t}+\kslash\eslash\}|e_L^-\rangle
\end{equation}
$$R_t = {m_t^2 \eta_1\over -2k\cdot p_{\bar b}(-2k\cdot p_{\bar t} + m_t^2 \eta_1)}
\left( F_\gamma^L F_{\gamma t\bar t}^L
{\cal R}_t^{(1)} - m_t^2F_\gamma^L F_{\gamma t\bar t}^R {\cal R}_t^{(2)}
\right)$$
$${\cal R}_t^{(1)} = \langle \bar b_R|\Wslash^-  (-\pslash_{\bar t}+\kslash)
| e_R^+\rangle
\langle e_L^-|\{(\pslash_t+\kslash)
\eslash\pslash_t+m_t^2\eslash\}\Wslash^{+\ast}|b_L\rangle$$
\begin{equation}
  \label{eq27}
{\cal R}_t^{(2)} = \langle e_R^+| \{2\epsilon\cdot p_t+\kslash\eslash\}
\Wslash^{+\ast} | b_L\rangle
\langle \bar b_R|\Wslash^-|e_L^-\rangle
\end{equation}
This set is gauge-invariant in the limit $\eta_2 = 0$ (if we set $\epsilon^\mu = k^\mu$, we
obtain ${\cal R}_{\bar b} + {\cal R}_{\bar t} + {\cal R}_t = 0$).
The soft and collinear limits are defined by $k^0 < \omega_0$ and $k_\perp / k^0 < x_0$, respectively,
where $\omega_0$ and $x_0$ are arbitrary small parameters. Let $\bar\Omega$ be the corresponding region
of the gluon phase space. The contribution to ${\cal A}_1$ is
\begin{equation}
  \label{eq28}
C_F (g\mu^\epsilon)^2
\int_{\bar\Omega} {d^{3-2\epsilon} k \over (2\pi)^{3-2\epsilon}k^0} \sum_{\epsilon_\pm} \left|
{\cal R}_{\bar b} + {\cal R}_{\bar t} + {\cal R}_t\right|^2
\end{equation}

\subsection{The t decay channel}

This channel can be obtained from the \=t decay channel by conjugation. Specifically, we define 
$p_{\bar t}^\mu = p_{W^-}^\mu + p_{\bar b}^\mu$ and
$p_t^\mu = p_{W^+}^\mu + p_b^\mu + k^\mu$ with $\eta_{1,2}$ given by Eq.~(\ref{eq7}).
We set $\eta_1 = 0$ in the matrix elements, so that
the integral over soft and collinear gluons contributes to ${\cal A}_2$ (Eq.~(\ref{eq12})).
We define the gluon polarization vector with respect to the b quark ({\em cf.}~ Eq.~(\ref{eq24})).
The phase space is defined by $|\eta_2 m_t^2| < |( p_t - k)^2 - m_t^2|$.
The amplitudes that contribute correspond to gluon emission from the \=t, t and b quarks (Fig.~\ref{fig3}{\em (b-d)}). This set is gauge
invariant in the limit $\eta_1 = 0$. One obtains expressions similar to Eqs.~(\ref{eq25}), (\ref{eq26})
and (\ref{eq27}).
The amplitude for real gluon emission from $b$ is
$$R_b = {1\over 2k\cdot p_b} \Big(F_\gamma^L F_{\gamma t\bar t}^L
{\cal R}_b^{(1)} - m^tF_\gamma^L F_{\gamma t\bar t}^R {\cal R}_b^{(2)}\Big)$$
$${\cal R}_b^{(1)} = \langle \bar b_R|\Wslash^-  \pslash_{\bar t}
| e_R^+\rangle
\langle e_L^-|\pslash_t\Wslash^{+\ast}
(\pslash_b+\kslash)\eslash|b_L\rangle$$
\begin{equation}
  \label{eq29}
{\cal R}_b^{(2)} = \langle e_R^+| \Wslash^{+\ast} (\pslash_b+\kslash)\eslash| b_L\rangle
\langle \bar b_R|\Wslash^-|e_L\rangle
\end{equation}
The amplitude for real gluon emission from t is
$$R_t = {1\over 2k\cdot p_t + m_t^2 \eta_2} \Big(F_\gamma^L F_{\gamma t\bar t}^L
{\cal R}_t^{(1)} - m_t^2F_\gamma^L F_{\gamma t\bar t}^R {\cal R}_t^{(2)}\Big)$$
$${\cal R}_t^{(1)} = \langle \bar b_R|\Wslash^-  \pslash_{\bar t}
| e_R^+\rangle
\langle e_L|\{\pslash_t
\eslash(\pslash_t-\kslash)+m_t^2\eslash\}\Wslash^{+\ast}|b_L\rangle$$
\begin{equation}
  \label{eq30}
{\cal R}_t^{(2)} = \langle e_R^+| \{2\epsilon\cdot p_t-\eslash\kslash)\}\Wslash^{+\ast} | b_L\rangle
\langle \bar b_R|\Wslash^-|e_L^-\rangle
\end{equation}
The amplitude for real gluon emission from \=t is
$$R_{\bar t} = {m_t^2 \eta_2 \over 2k\cdot p_b (2k\cdot p_t + m_t^2 \eta_2)} \Big(F_\gamma^L F_{\gamma t\bar t}^L
{\cal R}_{\bar t}^{(1)} - m_t^2F_\gamma^L F_{\gamma t\bar t}^R {\cal R}_{\bar t}^{(2)}\Big)$$
$${\cal R}_{\bar t}^{(1)} = \langle \bar b_R|\Wslash^- \{\pslash_{\bar t}
\eslash(\pslash_{\bar t}+\kslash)-m_t^2\eslash\}| e_R^+\rangle
\langle e_L|(\pslash_t-\kslash)\Wslash^{+\ast}|b_L\rangle$$
\begin{equation}
  \label{eq31}
{\cal R}_{\bar t}^{(2)} = \langle e_R^+| \Wslash^{+\ast} | b_L\rangle
\langle \bar b_R|\Wslash^-\{2\epsilon \cdot p_{\bar t}+\eslash\kslash\}|e-\rangle
\end{equation}
contributing to ${\cal A}_2$ in the soft ($k^0 < \omega_0$) and collinear ($k_\perp / k^0 < x_0$)
limits
\begin{equation}
  \label{eq32}
C_F (g\mu^\epsilon)^2\int_{\Omega} {d^{3-2\epsilon} k \over (2\pi)^{3-2\epsilon}k^0} \sum_{\epsilon_\pm} \left|
{\cal R}_{\bar b} + {\cal R}_{\bar t} + {\cal R}_t\right|^2
\end{equation}
where $\Omega$ is the region of gluon phase space in these limits.

\subsection{The t\=t production channel}

This channel contains contributions from gluon emission from all quark legs. Since we are only keeping terms
that are linear in $\eta_{1,2}$, we may divide the channel into two parts, corresponding to the two channels
we considered above. The $o(\eta_1)$ ($o(\eta_2)$) terms are given by the $o(\eta_1)$ ($o(\eta_2)$)
terms in the \=t (t) decay channel. The only difference is that in this channel we define
$p_{\bar t}^\mu = p_{W^-}^\mu + p_{\bar b}^\mu$ and
$p_t^\mu = p_{W^+}^\mu + p_b^\mu$. The phase space is restricted by
\begin{equation}
  \label{eq33}
|\eta_1 m_{\bar t}^2| < |( p_{\bar t} + k)^2 - m_t^2|\; ,\; |\eta_2 m_t^2| < |( p_t + k)^2 - m_t^2|
\end{equation}
In the limit $\eta_1 = \eta_2 = 0$, we obtain identical expressions from Eqs.~(\ref{eq25} - \ref{eq27})
and Eqs.~(\ref{eq29} - \ref{eq31}). In the soft limit ($k^0 < \omega_0$), we obtain
$${\cal A}^{soft} (0,0) = C_F (g\mu^\epsilon)^2
\int_{k^0 < \omega_0} {d^{3-2\epsilon} k \over (2\pi)^{3-2\epsilon}k^0}
\sum_{\epsilon_\pm} \left|
{\cal R}_{\bar t} + {\cal R}_t\right|^2$$
\begin{equation}
  \label{eq34}
= C_F (g\mu^\epsilon)^2
\int_{k^0 < \omega_0} {d^{3-2\epsilon} k \over (2\pi)^{3-2\epsilon}k^0}
\left( {p_{\bar t} \over 2k\cdot p_{\bar t}} - {p_t \over 2k\cdot p_t}\right)^2 {\cal T}_{\gamma , Z} (L,W^-,W^+)
\end{equation}
which is the correct infrared limit.

\section{The Monte Carlo event generator}
\label{sec4}

We used the formulas obtained above for the amplitudes to construct a Monte Carlo event generator.
The generator is based on a program written by Schmidt in C++. Schmidt's program generates events at tree
level with 100 \% efficiency. This is achieved by making use of helicity states, not only for the final
states, but also for the intermediate t and \=t quark states. The amplitude is then expressed in terms
of helicity angles which are defined in different Lorentz frames for the three vertices, respectively.
This construction is carried over to one-loop level, if one neglects gluon interference effects, because
the amplitude is still factorizable at that level. The virtual, soft and collinear corrections amount
to corrections in the form factors for the three vertices in the diagrams. The real gluon corrections
may also be grouped similarly into three well defined channels (t\=t production, \=t decay, and
t decay). This is because one may always identify on-shell t and \=t states in each diagram.
Thus, one can construct a very efficient event generator.

To account for gluon interference effects, one may not rely on helicity angles, because the amplitudes
are no longer factorizable, and need to be expressed in terms of a single Lorentz frame. This complicates their
calculation. To simplify the expressions, we have introduced polarization vectors for the intermediate
$W^\pm$ boson states defined with respect to the b (\=b) quarks (which are treated as massless).
This quadruples the number of diagrams, but simplifies the final expressions.
At tree level, we have checked the accuracy of our expressions numerically by using Schmidt's program.
At loop level, we have performed a similar check in the limit $\eta_{1,2} \to 0$.
Before we can feed our expressions including gluon interference effects to a Monte Carlo event generator,
we need to integrate over the momentum of the gluon (virtual, or real, accordingly).
In general, this can be done using the standard procedure implemented via the symbolic manipulation
program FORM~\cite{ref7}. In our case, we need to first reduce the number of gluon momentum instertions to two, otherwise
the final expressions turn out to be too lengthy to handle. This can easily be done if the amplitudes
are expressed in terms of $W$ boson polarization vectors, as described in the previous Section.
After the amplitudes have thus been massaged, they can be straightforwardly reduced to scalar amplitudes
via FORM~\cite{ref7}. The final expressions can then be fed to an event generator.

Since gluon interference effects are relatively small, we have taken advantage of Schmidt's efficient
code, by keeping the event generating procedure and introducing the new effects as a correction to the
weight. In general, this correction is an extra factor to the weight,
\begin{equation}
  \label{eq35}
  w_{new} = {{\cal A} (\eta_1, \eta_2)\over {\cal A} (0,0)} = 1 + {\cal A}_1 + {\cal A}_2
+ o(\eta_{1,2}^2)
\end{equation}
There is an additional complication in the case of hard gluon emission, where it is no longer possible
to define channels corresponding to the three main vertices in the diagrams (t\=t production, \=t decay, and
t decay) by identifying on-shell intermediate t and \=t quark states.
To maintain the structure of the program, we define the three channels by dividing the phase space of
the final states into distinct regions corresponding to the three channels, as described in the previous
Section. These definitions of the
channels agree with Schmidt's in the limit $\eta_{1,2} \to 0$.

In our approach, we have set the mass of the b quark to zero in matrix elements. This was essential for
the simplification of the expressions for the various amplitudes. We have also neglected effects of
$o(\eta_{1,2}^2)$, which include gluon interference between the b and \=b quarks. Such effects are
very small and are not likely to contribute at the desired accuracy for the NLC. Should the need for higher
accuracy arise, our method can be readily applied to accommodate these effects. We have also not coupled
the program with a jet algorithm. This is necessary for a direct comparison with experimental data.
Initial-state radiation effects~\cite{ref12} are included in the program, but will not be discussed here.

Next, we present numerical results obtained with the Monte Carlo event generator. As has been shown, there
is no correction to the total cross-section, but differential cross-sections are altered by gluon
interference effects. Our results include hard gluon effects and are in general agreement with the
semi-classical analysis of soft gluon production of Khoze, {\em et al.}~\cite{ref9}.
Specifically, we plot the distribution of gluon energy (Figs.~\ref{fig4},
\ref{fig5}), transverse momentum (Figs.~\ref{fig6}, \ref{fig7}), and
polar angle (Figs.~\ref{fig8}, \ref{fig9}).
All these plots are for center-of-mass energy $\sqrt s = 400$~GeV.
We have also introduced the jet resolution parameter $\eta_{cut} = 0.05$
by demanding
\begin{equation}
  \label{eq36}
(k+p_b)^2/s > \eta_{cut}^2 \;,\quad (k+p_{\bar b})^2/s > \eta_{cut}^2
\end{equation}
where $k^\mu$ is the four-momentum of the gluon.
We have imposed no other cut on the gluon.

In Figs.~\ref{fig4}, \ref{fig6}, and \ref{fig8}, we have compared distributions
with and without interference for the three variables we are studying.
In general, interference effects enter at a 10 \% level, so they need to be
included in a precision study of top quark production and decay.
Moreover, in Figs.~\ref{fig5}, \ref{fig7}, and \ref{fig9}, we show the
dependency of the respective distributions on the decay width of the top quark,
$\Gamma_t$. In Fig.~\ref{fig5}, we see that the decay width has a most
pronounced effect in the range of gluon energies $E_g \sim \Gamma_t$~\cite{ref6}, and similarly for the gluon transverse momentum (Fig.~\ref{fig7}).
Also, the effect of $\Gamma_t$ can be amplified in the distribution of
the gluon polar angle, if one tags the top or bottom quarks~\cite{ref6}.

\section{Conclusions}
\label{sec5}

We have performed a study of gluon interference effects in the t\=t production above threshold and the
subsequent semi-leptonic decay at the NLC. Our study extends previous analytical analyses of soft gluon
production, by including hard gluon effects. We have performed a numerical analysis which is complete
to $o(\Gamma_t/m_t)$. We have obtained analytical expressions for the amplitudes, aided by the
symboic manipulation routines based on FORM~\cite{ref7}. These expressions were then fed into a Monte Carlo
event generator based on code written by Schmidt.

Although there is no correction to the total cross-section, differential
cross-sections are affected by gluon interference at the 10\% level, in general. Therefore,
such effects need to be included in a precision analysis of top quark production and decay at the
NLC. A complete study should span the beam energy range including the region near threshold. This presents
the additional complication of color Coulomb effects which require non-perturbative methods to be dealt with.
The region near threshold has recently been studied in detail. We hope to extend the range of validity of
our Monte Carlo event generator by including Coulomb effects at threshold in the near future.

We have concentrated on the semi-leptonic decay channel, because this is where one obtains the
cleanest signal. The branching ratios for other (hadronic) channels are higher and gluon interference effects
are more pronounced there. We are planning to continue with a systematic study of the other top decay
channels. Our method can also be applied to hadronic colliders, such as the LHC, but a similar study in that
environment is far more involved, because of gluon interference between initial and final states.

\acknowledgments

I am indebted to Carl Schmidt for letting me use his program. I also wish to thank Linda Arvin and Eric
Conner for valuable help with the development of the program.

\newpage

%========================
\newpage
%========================
\newpage
\begin{figure}
 %
\vspace{6in}
\caption{t\=t production and decay with gluon emission.}
\label{fig3}
\end{figure}
\newpage
\begin{figure}
\begin{center}
%%%%%%%%%%%%%%%%%%%
% GNUPLOT: LaTeX picture
\setlength{\unitlength}{0.240900pt}
\ifx\plotpoint\undefined\newsavebox{\plotpoint}\fi
\sbox{\plotpoint}{\rule[-0.200pt]{0.400pt}{0.400pt}}%
%
\end{center}
\caption{Distribution of gluon energy with (dotted line)
and without (solid line) interference at center-of-mass energy of
400~GeV.}
\label{fig4}
\end{figure}
\begin{figure}
\begin{center}
%%%%%%%%%%%%%%%%%%
% GNUPLOT: LaTeX picture
\setlength{\unitlength}{0.240900pt}
\ifx\plotpoint\undefined\newsavebox{\plotpoint}\fi
\sbox{\plotpoint}{\rule[-0.200pt]{0.400pt}{0.400pt}}%
%
\end{center}
\caption{Distribution of gluon energy for $\Gamma_t = 0.7, 1.5, 10$~GeV
(including interference) at center-of-mass energy of 400~GeV.}
\label{fig5}
\end{figure}
\newpage
\begin{figure}
\begin{center}
%%%%%%%%%%%%%%%%%%%
% GNUPLOT: LaTeX picture
\setlength{\unitlength}{0.240900pt}
\ifx\plotpoint\undefined\newsavebox{\plotpoint}\fi
\sbox{\plotpoint}{\rule[-0.200pt]{0.400pt}{0.400pt}}%
%
\end{center}
\caption{Distribution of gluon transverse momentum with (dotted line)
and without (solid line) interference at center-of-mass energy of
400~GeV.}
\label{fig6}
\end{figure}
\begin{figure}
\begin{center}
%%%%%%%%%%%%%%%%%%%%%
% GNUPLOT: LaTeX picture
\setlength{\unitlength}{0.240900pt}
\ifx\plotpoint\undefined\newsavebox{\plotpoint}\fi
%
\end{center}
\caption{Distribution of gluon transverse momentum
for $\Gamma_t = 0.7, 1.5, 10$~GeV
(including interference) at center-of-mass energy of 400~GeV.}
\label{fig7}
\end{figure}
\newpage
\begin{figure}
\begin{center}
%%%%%%%%%%%%%%%%%%%%%
% GNUPLOT: LaTeX picture
\setlength{\unitlength}{0.240900pt}
\ifx\plotpoint\undefined\newsavebox{\plotpoint}\fi
%
\end{center}
\caption{Distribution of the gluon polar angle with (dotted line)
and without (solid line) interference at center-of-mass energy of
400~GeV.}
\label{fig8}
\end{figure}
\begin{figure}
\begin{center}
%%%%%%%%%%%%%%%%%%%%%
% GNUPLOT: LaTeX picture
\setlength{\unitlength}{0.240900pt}
\ifx\plotpoint\undefined\newsavebox{\plotpoint}\fi
\sbox{\plotpoint}{\rule[-0.200pt]{0.400pt}{0.400pt}}%
%
\end{center}
\caption{Distribution of the gluon polar angle
for $\Gamma_t = 0.7, 1.5, 10$~GeV
(including interference) at center-of-mass energy of 400~GeV.}
\label{fig9}
\end{figure}

%\end{document}
\end{document}